\documentstyle[aps,eqsecnum,pre,multicol,preprint]{revtex}
\tighten
\begin{document}
\title{\bf Effect of criticality on wetting layers: A Monte-Carlo simulation study}
\author{Nigel B. Wilding}
\address{Department of Physics and Astronomy, The University of Edinburgh,\\
Edinburgh EH9 3JZ, U.K.}

\author{Michael Krech} 
\address{Institut f\"{u}r Theoretische Physik, RWTH Aachen,\\D-52056 Aachen, Germany}

\input epsf

\maketitle

\begin{abstract} 

A solid substrate, when exposed to a vapour, can interact with it in
such a way that sufficiently close to liquid-vapour coexistence a
macroscopically thick liquid wetting layer is formed on the substrate
surface. If such a wetting transition occurs for a binary fluid mixture
in the vicinity of the critical end point of demixing transitions,
critical fluctuations lead to additional long-ranged interactions
(Casimir forces) within the wetting layer, changing its equilibrium
thickness. We demonstrate this effect by Monte-Carlo simulations of 
wetting layers of a symmetrical Lennard-Jones binary fluid mixture near
its critical end point. The results suggest that the effect should also
be detectable in corresponding wetting experiments.

\end{abstract}




\section{Introduction}

Simple liquids or binary liquid mixtures in unbounded space are
homogeneous and isotropic systems. In the vicinity of surfaces, 
however, these spatial symmetries are broken and the fluid phase
diagram exhibits new surface-induced features. Principal among these
are surface phase transitions such as wetting and drying which can
occur when a fluid comes into contact with a solid wall
\cite{SD88}. These transitions are brought about by
long-ranged dispersion (Van der Waals) interactions between the
particles of the fluid and those comprising the wall.

Long ranged interactions of a quite different origin occur when a {\em
critical} fluid is in contact with a wall \cite{KB83,HWD86}. 
Criticality is characterised by a divergent correlation length and
strong order parameter fluctuations. As a result of the long range
correlations, the wall modifies the properties of the critical fluid
over considerable distances and this can dramatically affect surface
properties such as wetting behaviour. Thus, for example, a binary
liquid mixture usually demixes close to a wall due to the preferential
affinity of the wall material for one of the two components
\cite{MEFHAY80}. If the demixing transition becomes critical, the
adsorbed amount of the preferred component becomes macroscopic, a
phenomenon known as critical adsorption \cite{HWDAC91,HWDMS93}. 

If the system is constrained in more than one spatial direction, e.g.
by the introduction of a second wall, the critical behaviour of the
fluid is modified again. When the correlation length becomes comparable
to the smallest linear extent of the system \cite{MEF70,MEFHN81,MNB83},
the size dependence of thermodynamic quantities is governed by
universal finite-size scaling functions. Such a finite geometry may be
imposed eg. when fluids are adsorbed in slits and pores, or generated
spontaneously in the course of a wetting transition eg. in a binary
liquid mixture near its critical end point of demixing
transitions \cite{SD88,MPNJOI85}. 

An important consequence of confinement in near-critical fluids is the 
generation of long-ranged forces between the confining walls
\cite{MKSD92,MK94}. This phenomenon is a direct analogue of the
well-known Casimir effect in electromagnetism \cite{LS96}. Formally,
the critical Casimir force arises as a result of the boundary
conditions placed on the spectrum of order parameter fluctuations by
the walls. This leads to a finite-size dependence for the singular part
of the critical free energy, and hence a separation-dependent force
between the walls \cite{MKSD92,MK94}. Such critical Casimir forces are
in addition to the usual dispersion forces which always exist in
confined systems  \cite{CCS94,GHNIB94,SLAR96,PJPA97,NSC92,BKM95}.
However, they differ from dispersion forces in the major respect that
they are governed by {\em universal} scaling functions
\cite{MNB83,MKSD92,MK94,MK97}. Precisely at the bulk critical point
these scaling functions take universal values known as Casimir
amplitudes \cite{MNB83,MKSD92,MK97}.

Much theoretical work has been directed in recent years towards an
understanding of the Casimir effect in critical systems. For the Ising
universality class, (which contains the critical demixing transition of
binary liquid mixtures), exact results are available in $d = 2$
dimensions for a strip geometry \cite{JLC86,REJS94}. In higher
dimensions, exact results are limited to the Casimir force scaling
function for the spherical model \cite{DD96}.  For the experimentally
relevant case of the $d=3$ Ising universality class, only approximate
results are available. These are based on real space renormalization
\cite{INW86}, the field theoretic renormalization group
\cite{MKSD92,MK97}, and Monte-Carlo (MC) simulations
\cite{MK97,MKDPL96} for the film geometry. Other geometries have also
been investigated with regard to the possibility of performing direct
surface force measurements of the Casimir force using atomic force
microscopy \cite{TWBEEUR95,SGUR95} (see also Ref.\cite{PJPA97}).
 
While the contemporary theoretical understanding of the Casimir effect
in critical fluids is rather advanced, the experimental situation is
not so satisfactory. Ideally, one should like to have experimental
estimates of Casimir amplitudes and scaling functions to compare with
theoretical estimates and to fill the gaps where no theoretical results
are available. To date, however, no direct experimental demonstrations
of the Casimir effect in critical fluid systems have been reported.
This state of affairs seems to be attributable to the difficulties of
performing high resolution surface force measurements within the
extremely tight constraints of reduced temperature and pressure
necessary to maintain criticality. In view of this, other more indirect
routes to investigating the Casimir force have been sought. Perhaps the
most promising of these is the effect of the Casimir force on wetting
layers of a binary fluid mixture. If the vapour of a binary fluid is
exposed to a solid substrate which strongly attracts the fluid
particles, a thick liquid wetting layer will form on the substrate for
state points sufficiently close to liquid-vapour coexistence. If, in
addition, the temperature is chosen close to the critical end point
(CEP) temperature (at which the line of critical demixing transitions
intersects the liquid-vapour coexistence curve), the presence of
Casimir forces within the wetting layer will change its equilibrium
thickness compared to a non-critical state point of the same
undersaturation \cite{MKSD92}. Since highly accurate measurements of
wetting film thicknesses are possible using modern ellipsometry
techniques, this would appear a rather promising approach to probing
the Casimir force experimentally. As yet, however, there have been no
clear reports of such an effect in experimental wetting studies of
binary fluid mixtures.

In view of the dearth of empirical results for the critical Casimir
force, we have undertaken a computer simulation investigation of
wetting behaviour close to the CEP of a binary fluid mixture. The aim
of our study was two-fold: on the one hand to investigate the
theoretical predictions of reference \cite{MKSD92}, and on the other to
try to make contact with real systems by working with as realistic a
simulation model as feasible.  Most previous simulation work on
critical phenomena in confined geometry has taken the form of MC
studies of lattice gas models, favoured for reasons of their
computational tractability \cite{KBDPL91,KBDPL92}. Such simulations are
usually performed using short-ranged surface fields to mimic the effect
of walls. In this work however, we have employed an off-lattice
Lennard-Jones (LJ) symmetrical binary fluid model at a planar solid
wall, and have incorporated the effects of long-ranged dispersion
interactions between the fluid and the wall. To our knowledge the
wetting behaviour of such a system has only been previously studied for
temperatures well below the consolute critical temperature
\cite{FFM93}. Hitherto, no attempts have been made to investigate
directly the effect of criticality on the formation of wetting layers
either in lattice or continuum fluid models.

Our paper is organised as follows. Section~\ref{sec:back} is devoted to
providing some brief background material concerning the bulk phase
diagram of symmetrical binary mixtures and the theory of the effect of
criticality on wetting layers.  In section~\ref{sec:mc} we introduce
our symmetrical binary fluid model and the MC simulation technique. We
then detail our simulation results for the wetting behaviour of the
mixture, and the influence of the critical point. Finally, in
section~\ref{sec:concs} we summarise and discuss our results.


\section{Background}  

\label{sec:back}
\subsection{Bulk phase behaviour of symmetrical binary mixtures}

\begin{figure}[b]
\setlength{\epsfxsize}{7.8cm}
\centerline{\mbox{\epsffile{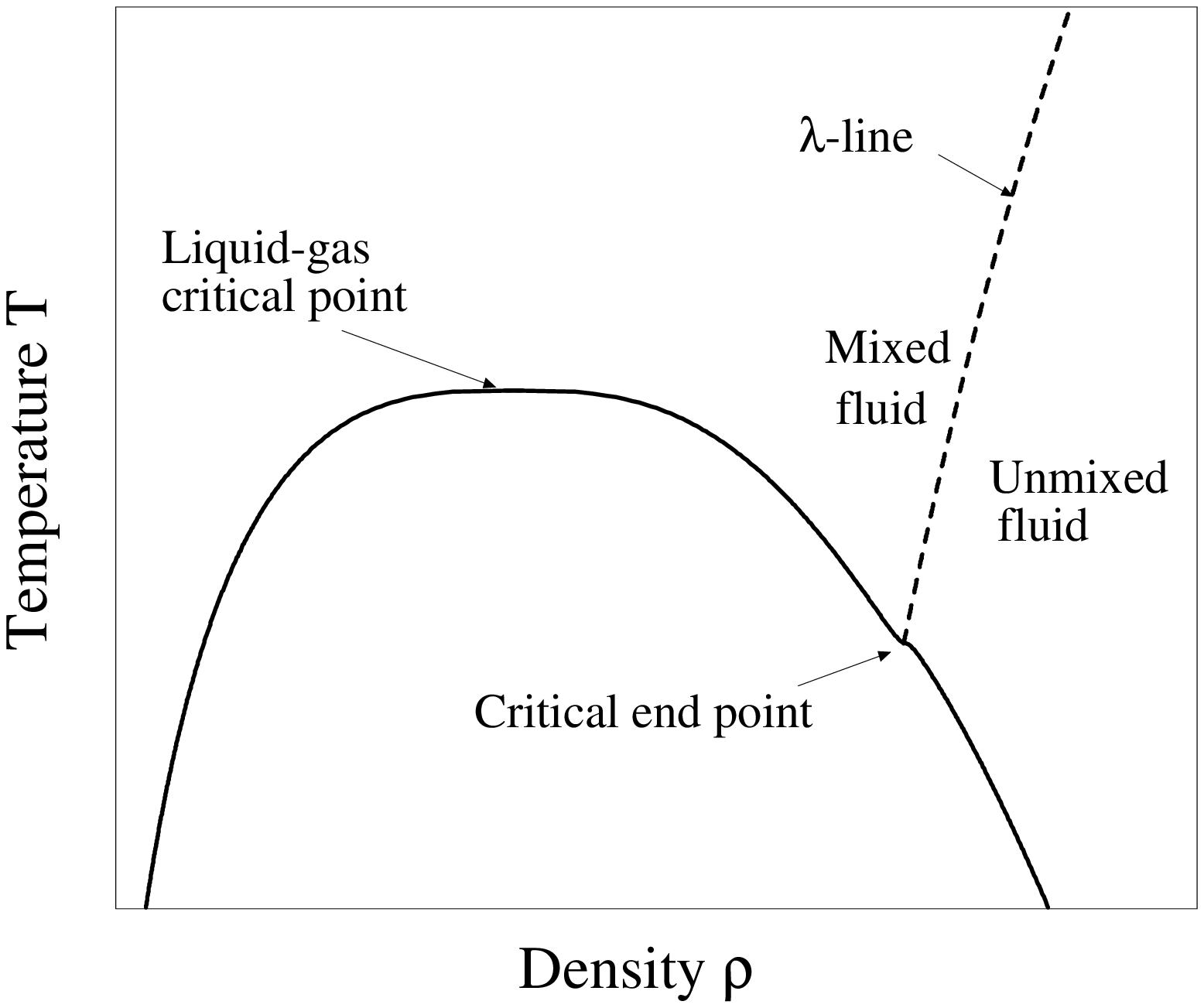}}} 

\caption{Schematic representation of the phase diagram of a symmetrical
binary fluid mixture in the density-temperature plane. The full curve
is the liquid-gas coexistence envelope. The dashed curve is the
$\lambda$-line of critical demixing transitions. The two curves
intersect at the critical end point. The singularity in the liquid
branch at the CEP is also shown \protect\cite{WILDING2}.}

\label{fig:mixschem}
\end{figure}

The simplest model for a binary fluid mixture is a symmetrical mixture
in which the pure components are identical, and only the interactions
between unlike particle species differ. Such a model can exhibit a variety
of phase diagram topologies depending on the relative strengths of the
two types of interactions. In this work, we shall be concerned with
symmetrical mixtures exhibiting a critical end point. The liquid-vapour
phase diagram of such a system in the density-temperature ($\rho-T$)
plane is shown schematically in figure~\ref{fig:mixschem}. At high
liquid densities there exists a `$\lambda$' line of consolute critical
points. Above this line the liquid is mixed, while below the line it is
demixed. As the liquid density is decreased, the demixing transition
temperature moves to lower temperatures. At some point, however, the
liquid phase becomes unstable with respect to the vapour. The
intersection between the $\lambda$ line and the liquid-gas coexistence
curve marks the critical end point, which is unique as being the only
point at which a critical liquid coexists with a non-critical gas.

Critical end points have been the subject of some interest in recent
years on account of the singularities they induce in the first-order
coexistence phase boundary (in this case the liquid-gas boundary).
These singularities have been studied both theoretically \cite{MFPU90}
and by simulation \cite{WILDING2}.  One is manifest as a bulge in the
liquid branch density at the CEP, as shown in
figure~\ref{fig:mixschem}. Another is found in the coexistence chemical
potential.  In the present work, however, we shall be concerned with
the influence of the CEP on the wetting properties at the liquid-gas
boundary and how it manifests the Casimir effect. In the following
subsection we review existing theoretical predictions for the Casimir
effect in wetting layers of a binary fluid mixture.

\subsection{Wetting layer thickness and criticality}

\begin{figure}[b]
\setlength{\epsfxsize}{8.0cm}
\centerline{\mbox{\epsffile{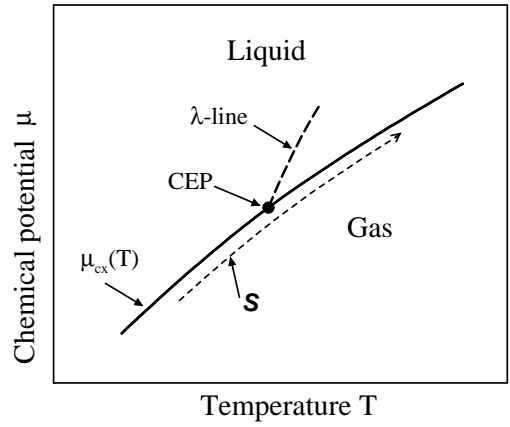}}} 

\caption{Schematic phase diagram of a symmetrical binary fluid.  The
full curve is the first order liquid-gas phase coexistence line
$\mu_{cx}(T)$.  The dashed curve is the critical ($\lambda$) line of
second order transitions separating the mixed and demixed liquid
phases.  The two curves intersect at the critical end point. Also
shown is a path ${\cal S}$ parallel to $\mu_{cx}(T)$ on the gas side.}

\label{fig:mixschem1}
\end{figure}

To analyse theoretically the wetting behaviour of the symmetrical
binary fluid in the vicinity of the CEP, it is expedient to consider
the phase diagram in the chemical potential-temperature ($\mu-T$)
plane, as shown schematically in figure~\ref{fig:mixschem1}. Indicated
on this diagram are the liquid-vapour coexistence line $\mu_{cx}(T)$,
the $\lambda$ line, and the CEP.

Consider now the thermodynamic path labelled ${\cal S}$ in
figure~\ref{fig:mixschem1}, which is given by the expression

\begin{equation}
\mu_{\cal S}(T)=\mu_{cx}(T)-\delta\mu,
\end{equation}
where $\delta\mu>0$ is a constant. This path runs parallel to the
liquid-vapour coexistence curve and passes the critical end point on
the vapour side. In the presence of a sufficiently attractive wall
(complete wetting regime), and for sufficiently small undersaturations
(i.e. $\delta\mu$ small), a thick liquid wetting layer will form at
the wall for all points on ${\cal S}$. An expression for the thickness
of such a wetting layer is obtainable by considering the layer free
energy per unit area (effective interface potential) as a function of
its thickness $l$. In the limit of an infinite wall, this is given by
\cite{MKSD92}

\begin{equation} 
\omega(l)=l(\rho_l-\rho_v)\delta\mu+\sigma_{wl}+\sigma_{lv}+\delta\omega(l) 
\label{eq:eip}
\end{equation}
Here the first term on the right hand side represents the free energy
penalty of building up a liquid layer when the vapour is the
stable bulk phase. The surface tension terms $\sigma_{wl}$ and
$\sigma_{lv}$ contain the effects of short ranged interactions at the
wall-liquid and liquid-vapour interfaces respectively, and are independent of $l$.
The fourth term,  $\delta\omega(l)$, is a finite size term incorporating
both the long ranged dispersion forces and any critical finite-size
effects. 

Precisely at criticality, and for an infinite wall,  $\delta\omega(l)$
takes the form \cite{MKSD92}:

\begin{equation} 
\delta\omega(l)\simeq\frac{W}{l^2}+\frac{k_BT\Delta}{l^2}.
\label{eq:omfs} 
\end{equation}
In this expression, the first term on the right hand side represents
the finite-size dependence of the (non retarded) dispersion forces
\cite{PdG81}, whose amplitude is given by the Hamaker constant $W$. The
second term represents the critical finite-size contribution to the
free energy \cite{MFPdG78} in three dimensions, controlled by a
universal Casimir amplitude $\Delta$, the magnitude and sign of which
depend on the nature of the boundary conditions on the wetting layer.
For $\Delta$ positive, the Casimir force is repulsive leading to a
layer thickening, while for $\Delta$ negative, it is attractive
resulting in thinning of the layer. The $\simeq$ in eq.~\ref{eq:omfs}
indicates that we neglect any density gradient in the liquid layer that
may come about because of the attractive wall potential. A density
gradient implies that not all the liquid layer is precisely at
criticality. However, this is expected to result in relatively
small corrections to eq.~\ref{eq:omfs}.

Minimising $\omega(l)$ with respect to $l$ yields the equilibrium layer
thickness $L_c$ at criticality:

\begin{equation}
L_c\simeq\left ( \frac{2k_BT\Delta+2W}{\delta\mu(\rho_l-\rho_v)}\right )^{1/3}.
\end{equation}
For points on ${\cal S}$ away from the CEP, however, the critical finite-size term
in eq.~\ref{eq:omfs} drops out. A useful quantity is thus the
ratio of the equilibrium thickness at criticality $L_c$ to its value $L_0$
away from criticality, for which one finds \cite{MK94}

\begin{equation} 
\frac{L_c}{L_0}\simeq\left (1+k_BT_{c}\frac{\Delta}{W}\right )^{1/3},
\label{eq:Lchg}
\end{equation} 
to leading order in $\delta\mu$. Here we have ignored any
temperature dependence of the Hamaker constant.

This relationship expresses the change in thickness of a wetting layer
as it becomes critical, and relates it to the Casimir amplitude
$\Delta$. As such it provides a potentially sensitive probe of the
Casimir effect itself.

\section{Monte Carlo studies}
\subsection{Model and simulation details}
\label{sec:mc}

The system we have studied is a symmetrical binary fluid,
having interparticle interactions of the Lennard-Jones (LJ) form:

\begin{equation}
u(r_{ij})=4\epsilon_{ij}\left[\left(\frac{\sigma_{ij}}{r_{ij}}\right)^{12}-\left(\frac{\sigma_{ij}}{r_{ij}}\right)^6\right]
\end{equation} 
The following choice of parameters was made: $\sigma_{11}= \sigma_{22}=
\sigma_{12}=\sigma=1$; $\epsilon_{11}= \epsilon_{22}= \epsilon$;
$\epsilon_{12}= 0.7\epsilon$. i.e. the pure components are identical,
but the unlike interactions are weakened. In common with most previous
simulations of LJ systems, the inter-particle potential was truncated
at a distance of $R_c=2.5\sigma$. No long-range correction or
potential shift was applied.

The fluid was confined within a cuboidal simulation cell having
dimensions $P_x\times P_y\times D$, in the $x,y$ and $z$ coordinate
directions respectively, with $P_x=P_y\equiv P$. As in previous work on
LJ fluids\cite{WILDING1}, the simulation cell was divided into cubic
sub-cells (of size the cutoff $R_c$) in order to aid identification of
particle interactions. Thus  $P=pR_c$  and $D=dR_c$, with $p$ and $d$
both integers.  To approximate a slit geometry, periodic boundary
conditions were applied in the $x$ and $y$ directions, while hard walls
were applied in the unique $z$ direction at $z=0$ and $z=D$. The hard
wall at $z=0$ was made attractive, using a potential designed to mimic
the long-ranged dispersion forces between the wall and the fluid
\cite{ISRAEL}:

\begin{equation}
V(z)=\epsilon_w\left[\frac{2}{15}\left(\frac{\sigma_w}{z}\right)^9-
\left(\frac{\sigma_w}{z}\right)^3\right]
\end{equation}
Here $z$ measures the perpendicular distance from the wall,
$\epsilon_w$ is a `well-depth' controlling the interaction strength,
and we set $\sigma_w=1$. No cutoff was employed and the wall potential
was made to act {\em equally} on both particle species.

Monte-Carlo simulations were performed using a Metropolis algorithm
within the grand canonical ($\mu,V,T$) ensemble \cite{AT87,WILDING1}. Three
types of Monte-Carlo moves were employed: 

\begin{enumerate}
\item Particle displacements
\item Particle insertions and deletions
\item Particle identity swaps: $1\to 2$ or $2\to 1$
\end{enumerate}

By virtue of the symmetry of the system, the chemical potentials
$\mu_1$ and $\mu_2$ of the two components were set equal at all times.
Thus only one free parameter, $\mu=\mu_1=\mu_2$, couples to the density.
The other variables used to explore the wetting phase diagram were the
reduced well depth $\epsilon/k_BT$ and the reduced wall potential
$\epsilon_w/k_BT$. During the simulations, the principal observable
monitored was the total particle density profile:

\begin{equation} 
\rho(z)= [N_1(z)+N_2(z)]/(P^2)
\end{equation} 
which was accumulated in the form of a histogram. Here $N_1$ and $N_2$
are the number of particles of the respective species. Other observables
monitored were the total interparticle energy and the wall interaction
energy.

The wetting layer thickness was obtained from the density profile as
$L=\rho_l^{-1}\int\rho(z)dz$ where $\rho_l$ is the liquid density.
Although more sophisticated definitions of absolute layer thickness can
be envisaged, this one is adequate for our purposes of detecting
relative thickness changes arising from Casimir forces.

\subsection{Method and results}

\label{sec:res}

The bulk liquid-vapour coexistence curve properties of the symmetrical
LJ fluid (with $\epsilon_{12}=0.7\epsilon$) have recently been
investigated by MC simulation \cite{WILDING2}. This work employed
state-of-the art simulation techniques to investigate the nature of
coexistence curve singularities induced by the critical end point. In
the course of the study, highly accurate estimates were obtained for
the location of the critical end point. Additionally, the locus of the
liquid-gas coexistence curve $\mu_{cx}(T)$ was measured to $5$
significant figures in the neighbourhood of the CEP.

Armed with accurate estimates for $\mu_{cx}(T)$, one is in a position
to perform detailed wetting studies very close to coexistence. Detailed
knowledge of $\mu_{cx}(T)$ is a prerequisite for obtaining a thick
wetting layer and detecting changes in its thickness due to Casimir
forces. Below we discuss the procedure we have employed for detecting
such changes.

Clearly for a liquid wetting layer to display critical behaviour at the
bulk CEP, it must be sufficiently thick to exhibit quasi-3D properties.
But for a thick wetting layer to form at a wall at all, the attractive
wall potential must be sufficiently strong that complete wetting occurs
at coexistence. To ensure that the CEP lay well within the complete
wetting regime, a number of preliminary test runs were performed in
which the temperature was set to its CEP value ($T_{cep}=0.958$) and
the density profile $\rho(z)$ monitored as coexistence was approached
in a sequence of steps from the gas side. To achieve  this, the
chemical potential was simply increased towards its coexistence value
in small increments of size $\Delta\mu=0.0025$. The procedure was repeated
for a number of different values of the wall potential $\epsilon_w$.

\begin{figure}[t]
\setlength{\epsfxsize}{8.0cm}
\centerline{\mbox{\epsffile{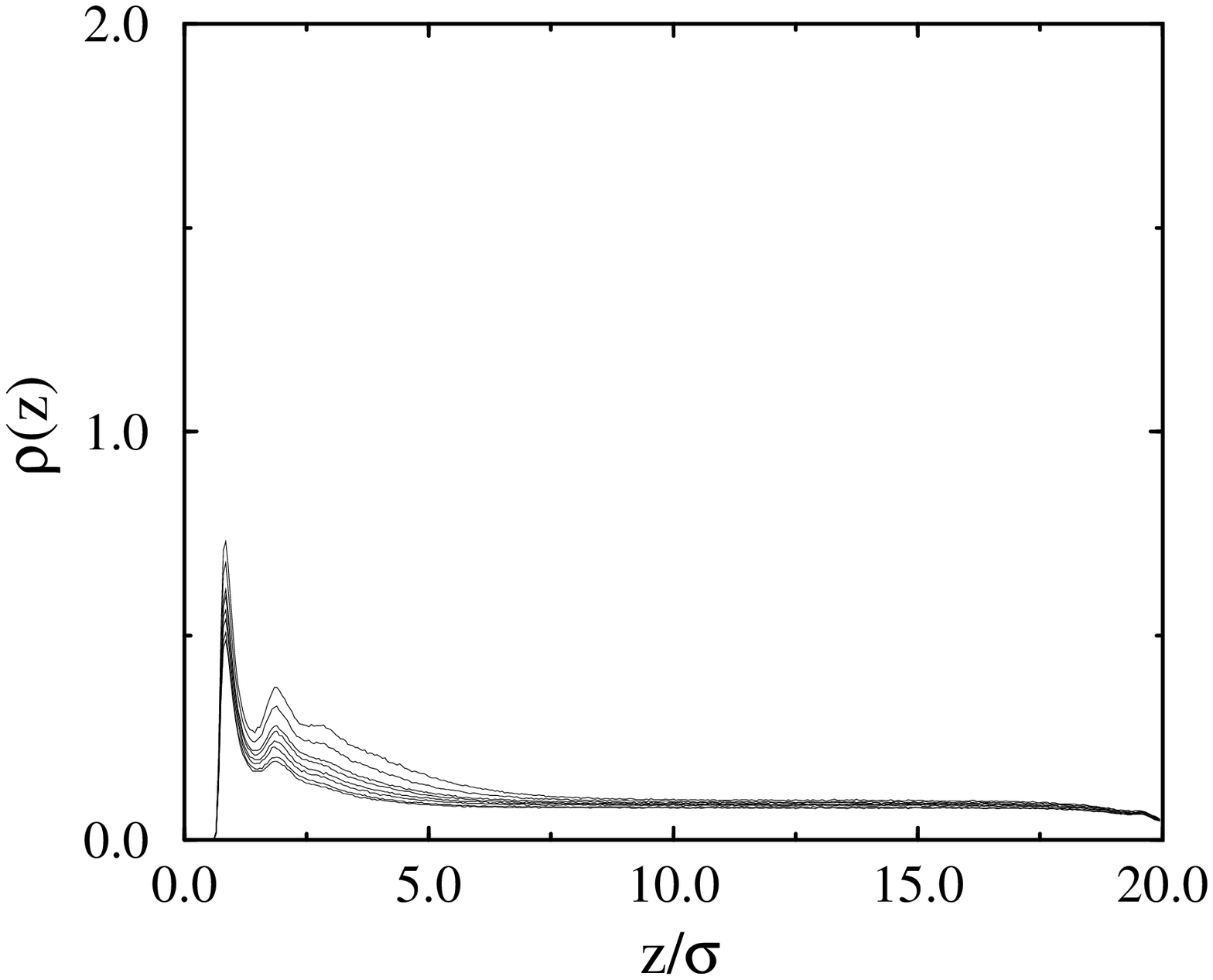}}} 
\centerline{\mbox{\epsffile{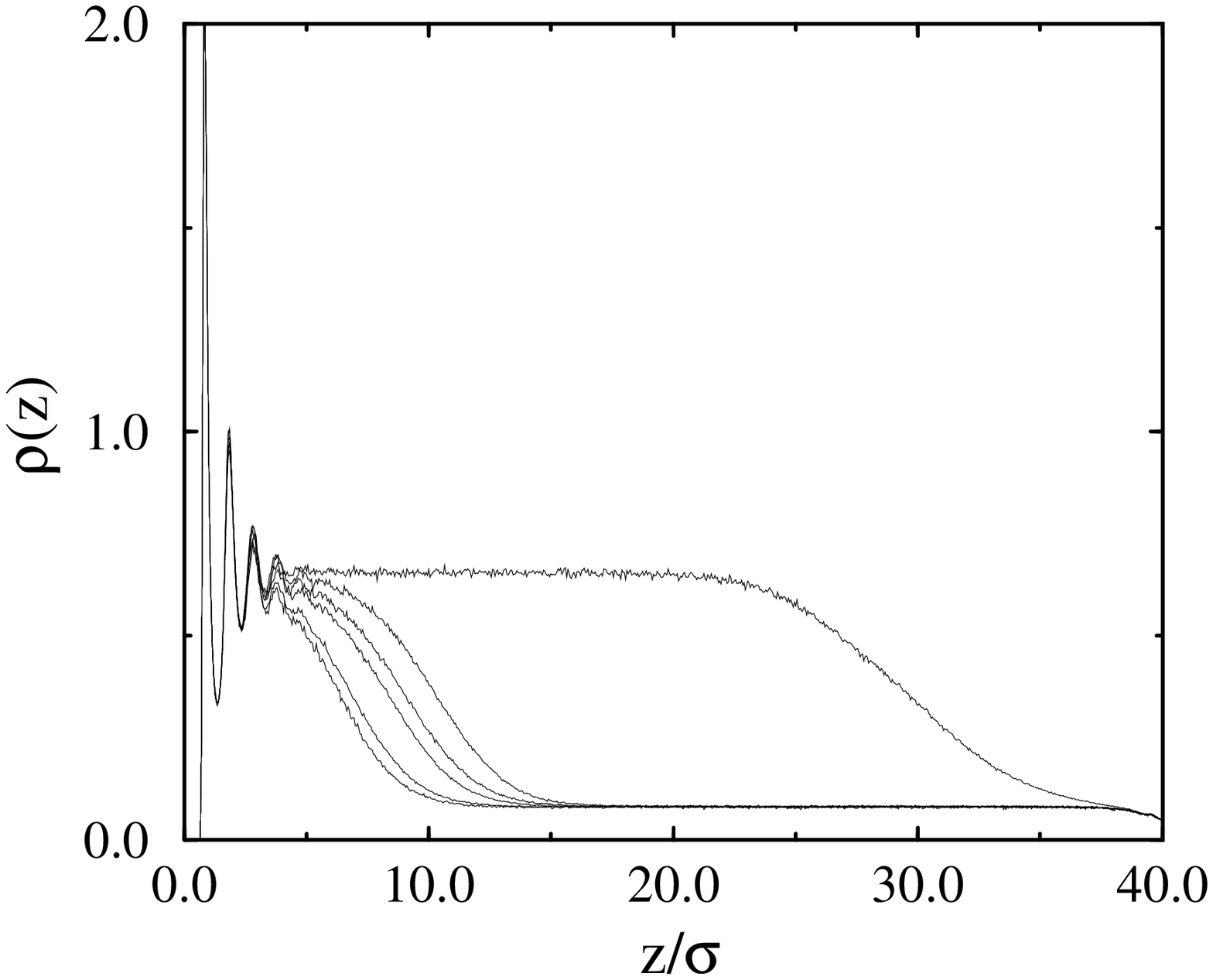}}} 

\caption{{\bf (a)} Density profiles $\rho(z)$ on the approach to
coexistence for $\epsilon_w=1.7$. The film thickness remains small
right up to coexistence signifying incomplete wetting. {\bf (b)}
Density profiles  for $\epsilon_w=1.8$. The film thickness grows very
large as coexistence is approached, signifying complete wetting.}

\label{fig:wet}
\end{figure}

For small values of $\epsilon_w/k_BT<1.7$, it was found that the film
thickness grew as coexistence was approached, but always remained
finite (cf. figure~\ref{fig:wet}(a)). The presence of a thin wetting
layer right up to coexistence signals incomplete (partial) wetting. For
stronger wall potentials ($\epsilon_w/k_BT>1.8$), however, the film
thickness became very large as coexistence was approached and
ultimately almost completely filled the system (cf.
figure~\ref{fig:wet}(b)). This suggests that the wetting transition
lies in the range $1.7<\epsilon_w/k_BT<1.8$. Nevertheless, to be quite
sure that the CEP lay well above the wetting transition (ie. well
within the complete wetting regime), all subsequent work employed
$\epsilon_w/k_BT=3.0$ at the CEP.

We have studied the layer thickness $L(\mu_{\cal S}(T))$ along the path
$\mu_{\cal S}(T)=\mu_{cx}(T)-\delta\mu$. Given $\mu_{cx}(T)$, this path
is specified solely by the value of $\delta\mu$, which has to be
chosen as a compromise between two competing factors. On the one hand, 
$\delta\mu$ must be sufficiently small that a thick wetting layer
forms. The thickness of this layer must be sufficiently large that
quasi-3D critical behaviour occurs, and that the liquid-gas interfacial
properties are not unduly influenced by short-ranged packing effect
close to the attractive wall. On the other hand, one wishes to avoid
having a layer that is very thick, lest the computational expense get
out of hand. A minimum film thickness of $L=10\sigma$ was judged
sufficiently large for our purposes. However, since close to
coexistence the layer thickness is extremely sensitive to changes in
$\delta\mu$, fine tuning was necessary. It turned out that to obtain
the required minimum thickness necessitated use of an extremely small
$\delta\mu$, namely $\delta\mu=0.002$. (corresponding to a relative
undersaturation $\delta\mu/\mu\sim10^{-3}$). At this undersaturation,
$L$ was observed to fluctuate in the range $10\sigma\lesssim L\lesssim 13\sigma$. 

Another condition to be satisfied in the simulations, is that the
largest fluctuation in the layer thickness be small compared to the
linear extent of the system in the $z$ direction, ie. $L\ll D$. This
latter condition ensures that the liquid gas interface does not
interact with the hard-wall at $z=D$, which might otherwise obscure the
effects of criticality, or even lead to capillary condensation.
Accordingly the choice $D=40\sigma$ was made.

Ideally, in order to facilitate direct contact with real systems and
existing theoretical work, one should like to simulate an effectively
semi-infinite system having lateral dimensions $P\gg L$. In
this limit, the finite-size critical  behaviour of the system depends
only on the ratio $\xi/L$ of the correlation length to the layer
thickness, and not on the lateral dimensions of the simulation box
\cite{MFPdG78}. Unfortunately, in this work we could not approach the
semi-infinite limit as closely as we would have liked because of the
huge computational expense entailed. In fact the largest lateral
dimensions that we could reach were $P=12.5\sigma,
P=15\sigma$ and $P=17.5\sigma$, containing average particle
numbers of approximately $N=1500, 1800, 3500$ respectively. We postpone
discussion of the effects of the relative smallness of $P$ 
until later.

The limiting factor in the speed of the simulations and (hence the
attainable system sizes) was found to be the extreme slowness of the
interfacial fluctuations, which rendered it very difficult to
accumulate accurate estimates of the average thickness.  In an effort
to ameliorate this problem, multi-histogram reweighting techniques were
employed \cite{AFRS88}. These allow one to combine data accumulated at
individual state points and, by extrapolation, obtain estimates for
observables at other not-too-distant state points. In our study the
computational complexity meant that it was feasible to accumulate
$\rho(z)$ at just three points on the path ${\cal S}$ for each system
size (see figure~\ref{fig:mixschem1}. These simulation points were
chosen to span the CEP temperature, namely $T=0.946, 0.958, 0.97$.
Subsequently, the data from the individual simulations were combined
self-consistently and extrapolated to yield estimates for $\rho(z)$
along the whole path. The data were collected from runs comprising in
total approximately $3\times10^9$  Monte Carlo steps (MCS), where we
define a MCS to be an attempt to perform each of the three types of MC
move: particle displacement, insertion/deletion, and identity swap (cf.
sec~\ref{sec:mc}). 

\begin{figure}[h]
\setlength{\epsfxsize}{8.0cm}
\centerline{\mbox{\epsffile{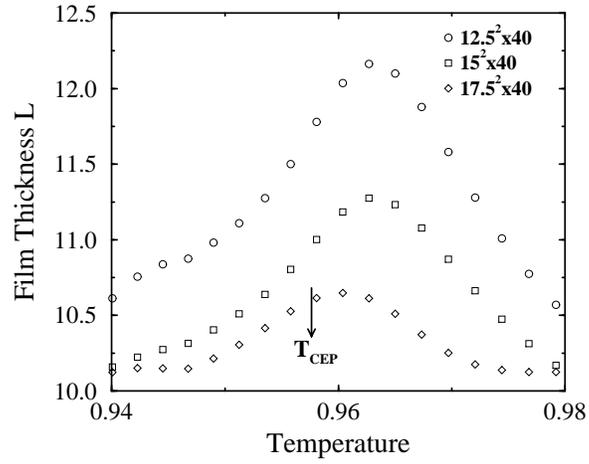}}} 

\caption{The thickness of the wetting layer as a function of
temperature along the thermodynamic path ${\cal S}$ defined in the text.
Data is shown for each of the three system sizes studied. The results
were obtained from multihistogram extrapolation of simulation data
accumulated at three points on this line, corresponding to temperatures
$T=0.946, 0.958, 0.97$.} 

\label{fig:thick}
\end{figure}

In figure~\ref{fig:thick} we present our results for the layer
thickness $L(\mu_{\cal S}(T))$, obtained by implementing this
procedure. The main feature of the results for each system size 
$P$, is a peak in the film thickness close to the CEP. For the
largest system size, the thickness either side of the peak is fairly
constant, while for the smallest system size, the accessible range of
temperature was not sufficient to encompass the whole peak. For the
smallest system size, the critical thickening is $\gtrsim 10\%$, while
for the largest system size it is $\approx 5\%$. The peak position is
at higher temperatures than the bulk CEP temperature and is closer to
the CEP for the largest system size than for the smallest. The peak
width also clearly narrows with increasing system size. We should
caution, however, that the statistical quality of our data is not
particularly high due to the difficulties mentioned above in collecting
statistics. The smoothness of the temperature dependence for
$L(\mu_{\cal S}(T))$ is to some extent an artifact of the histogram
extrapolation procedure and we are not particularly confident of our
estimates for the absolute film thickness. Nevertheless, we {\em are}
confident of the ability of our procedure to identify {\em changes} in
the wetting layer thickness.

The results for the film thickening due to criticality can be compared
with theoretical predictions for the semi-infinite system. In this
limit, the change in layer thickness is given by
eq.~\ref{eq:Lchg}. For our system, we calculate the Hamaker
constant to be $W\approx 2.5$ at the CEP. The value of the Casimir
amplitude $\Delta$ depends on the boundary conditions on the wetting 
layer. In the present model, these are of the form $(+,0)$ in the
notation of reference \cite{MK94}. This notation denotes an order
parameter (concentration) that is pinned to a constant value at the
wall due to the high particle density there, and which vanishes (on
average) at the liquid-gas interface because of the low gas density.
The most recent field theoretical and MC estimates \cite{MK97} yield
$\Delta(+,0)\approx 0.2$. Inserting this into eq.~\ref{eq:Lchg}
gives $L_c/L_0\approx 1.025$. Clearly this is a smaller thickening than
we see, but the trend of our results is in the direction of this value
and could lie close to the prediction for sufficiently large system
size.

In fact, it is possible to use finite-size scaling arguments to account
for the direction of the observed trend in peak height with increasing
$P$. To this end, let us rewrite the finite-size part of the critical
effective interface potential~\ref{eq:omfs}, taking account of the
finite lateral extent of the system

\begin{equation}
\delta\omega(l)=\frac{W}{l^2}+\frac{k_BT_c\Delta(+,0)}{l^2}+\frac{2l\Delta_{per}}{P^3}
\end{equation}

Here we have simply added an additional scaling term to account for the
finite system size $P$ in the periodic $x$ and $y$ directions. The
alteration to the free energy is assumed to scale like $l/P$, so that
the contribution to $\omega(l)$ (free energy per unit area) scales like
$l/P^3$. This new term can be interpreted as a finite-size correction
to the chemical potential.

Minimising this expression as before, gives the equilibrium critical thickness
$L_c$. The ratio of critical to non-critical thickness then follows as

\begin{equation}
\frac{L_c}{L_0}=\left(1+\frac{k_BT_c\Delta(+,0)}{W}\right )^{1/3}\left
[1+\frac{2k_BT_c\Delta_{per}}{\delta\mu(\rho_l-\rho_v)P^3}\right ]^{-1/3}
\end{equation}

The Casimir amplitude for periodic boundaries has been estimated as
$\Delta_{per}\approx -0.15$ \cite{MKDPL96}. Thus the correction factor
in $P$ is larger than unity and decreases towards unity as $P$
increases. If we take the liquid-gas density difference
$\rho_l-\rho_g\approx 0.5$, one finds that the correction factor is
approximately $1.06$ for the $P=12.5\sigma$ system size, reducing to
approximately $1.02$ for the $P=17.5\sigma$ system size. This accounts
semi-quantitatively for the observed decrease in the size of the
critical thickening as we increase $P$.

The influence of the periodic boundaries is presumably also responsible
for the the observed position of the peak in $L(\mu_{\cal S}(T))$. For
a semi-infinite system the finite-size scaling function for $\omega(T)$
displays a peak for $T<T_{cep}$ and thus the peak in $L(\mu_{\cal
S}(T))$ is also expected to occur for $T<T_{cep}$. In our work,
however, the peak in $L(\mu_{\cal S}(T))$ occurs for $T>T_{cep}$. It
seems reasonable, though, that were one to employ the finite-size
scaling function for periodic boundary conditions in $\omega(T)$ then
this could actually change the temperature dependence of $L(\mu_{\cal
S}(T))$ such that the maximum appears at some $T>T_{cep}$, if $P$ is
not much greater than $L$. Unfortunately, the scaling function for a
system with two periodic and one symmetry breaking boundary conditions
is presently not known to any approximation.


\section{Conclusions and outlook} 
\label{sec:concs}

In summary we have performed extensive Monte-Carlo simulations of
wetting behaviour near the critical end point of a symmetrical binary
fluid in contact with an attractive hard wall. The results demonstrate
that critical fluctuations within the wetting layer engender Casimir
forces, leading to small but significant changes in the layer thickness.
As such, they constitute the first empirical evidence for the Casimir
effect in a critical system.  Although an unambiguous determination of
the limiting size of the critical thickening was hindered by large
finite-size effects, the results are in order-of-magnitude agreement
with theoretical predictions for a semi-infinite system. Indeed, a
theoretical analysis of finite-size corrections to this scaling limit
yields estimated size effects comparable to those observed in the
simulations. Given the likely magnitude of the thickening in the
semi-infinite limit (which we estimate to be $\sim 3\%$), it seems
likely that the effect should also be experimentally detectable in real
(ie. non-symmetrical) absorbed binary fluids, many of which exhibit
Hamaker constants of similar magnitude to the present model.

With regard to the computational issues raised by our study,  our
results testify to the capability of modern simulation techniques to
identify small changes in the wetting behaviour of realistic fluid
models. Indeed, simulation is likely to prove of increasing value in
the study of wetting by critical layers, since unlike field theoretical
or density functional methods, it allows one both to tackle realistic
systems and to deal properly with critical fluctuations. Nevertheless,
algorithmic improvements are clearly desirable and necessary if  one is
to realise geometries that are effectively semi-infinite. The chief
problem experienced in this work was of extremely slow fluctuations of
the liquid-gas interface. In view of this, it may well pay dividends in
future work to employ a simulation algorithm that focuses more of the
computational effort on the interfacial region itself, this being the
bottleneck for phase space evolution. Such algorithmic improvements are
the matter of ongoing work. 

Finally, looking ahead to future work, it would doubtless be
interesting to investigate the nature of the interfacial properties
between the critical liquid and non-critical gas phases. This matter
has already been the subject of a number of detailed theoretical
investigations aimed at elucidating the temperature dependence of the
interfacial shape and surface tension on the approach to the CEP
\cite{WID72,TGREIH84,MFPU90}. Given present capabilities, a simulation
study of interfacial properties at the CEP would certainly seem
feasible and would nicely complement existing experimental studies
\cite{BL91,BL93}.

\acknowledgements

The authors are grateful to R. Evans for useful correspondence and to
M.E. Fisher for a careful reading of the manuscript. NBW thanks K.
Binder and  V. Privman for helpful discussions, and the Royal Society
of Edinburgh for financial support. MK gratefully acknowledges
financial support though the Heisenberg program of the Deutsche
Forschungsgemeinschaft.



\newpage

\newpage

\newpage

\end{document}